\documentclass{styles/vgtc}                          

\ifpdf
  \pdfoutput=1\relax                   
  \usepackage{graphicx}                
  \DeclareGraphicsExtensions{.pdf,.png,.jpg,.jpeg} 
\else
  \usepackage{graphicx}                
  \DeclareGraphicsExtensions{.eps}     
\fi%

\graphicspath{{figures/}{pictures/}{images/}{./}} 

\usepackage{microtype}                 
\PassOptionsToPackage{warn}{textcomp}  
\usepackage{textcomp}                  
\usepackage{mathptmx}                  
\usepackage{times}                     
\usepackage{cite}                      
\usepackage{tabu}                      
\usepackage{booktabs}                  

\newcommand{\etal}{{et al.\ }}

\usepackage{booktabs}
\usepackage{enumitem}
\setlist{leftmargin=1cm}

\usepackage{review}

\usepackage{color}
\definecolor{cb_green}{rgb}{0.3, 0.67, 0.29}
\definecolor{cb_blue}{rgb}{0.22, 0.49, 0.72}
\definecolor{cb_purple}{rgb}{0.59, 0.3, 0.64}

\onlineid{102}

\vgtccategory{Research}

\vgtcinsertpkg

\title{Worksheets for Guiding Novices through the Visualization Design Process}

\author{Sean McKenna\thanks{e-mail: sean@cs.utah.edu}\\
  \scriptsize University of Utah
\and Alexander Lex\thanks{e-mail: alex@sci.utah.edu}\\
  \scriptsize University of Utah
\and Miriah Meyer\thanks{e-mail: miriah@cs.utah.edu}\\
  \scriptsize University of Utah
}

\abstract{
  For visualization pedagogy, an important but challenging notion to teach is
  design, from making to evaluating visualization encodings, user interactions,
  or data visualization systems. In our previous work, we introduced the design
  activity framework to codify the high-level activities of the visualization
  design process. This framework has helped structure experts' design processes
  to create visualization systems, but the framework's four activities lack a
  breakdown into steps with a concrete example to help novices utilizing this
  framework in their own real-world design process. To provide students with
  such concrete guidelines, we created worksheets for each design activity:
  \textit{understand}, \textit{ideate}, \textit{make}, and \textit{deploy}. Each
  worksheet presents a high-level summary of the activity with actionable,
  guided steps for a novice designer to follow. We validated the use of this
  framework and the worksheets in a graduate-level visualization course taught
  at our university. For this evaluation, we surveyed the class and conducted 13
  student interviews to garner qualitative, open-ended feedback and suggestions
  on the worksheets. We conclude this work with a discussion and highlight
  various areas for future work on improving visualization design pedagogy.
}

\CCScatlist{ 
  \CCScat{H.5.2}{Information Interfaces and Presentation}{User Interfaces}{User-centered design}
}

\begin{document}

\maketitle

\section{Introduction}

To teach design in data visualization, educators combine many foundational
components, from user interface principles~\cite{Shneiderman2004} to data and
encoding taxonomies~\cite{Munzner2014}. Additional pedagogical materials for the
field focus on visual or perceptual
principles~\cite{Tufte1986,cairo2012,Ware2010} as a basis for creating and
judging data visualizations. Educators may also apply these principles and
techniques in the classroom through the use of design critiques or a cumulative
project. These visualization projects can be guided by several textbooks that
expound upon different design processes~\cite{Ware2010,Shneiderman2004} or
design decision models~\cite{Munzner2014} to help novice visualization designers
learn how to effectively and methodically build and validate visualization
systems.

Many of the pedagogical approaches to the visualization design process, however,
are theoretical in nature. From our own combined teaching experiences across 13
courses, we have witnessed students struggle to incorporate these theory-based
design concepts in practical, hands-on projects. As such, we believe there is a
need for new approaches to teaching the next generation of visualization
designers, equipping them with not just theoretical knowledge but also the
practical steps for building better systems and tools.

In our previous work~\cite{McKenna2014}, we introduced one such theoretical
model of the design process with four design activities: \textit{understand},
\textit{ideate}, \textit{make}, and \textit{deploy}. Each activity includes a
goal, produces a visualization artifact, and provides a plethora of design
methods to choose from. We found, however, that the theoretical framing of the
model restricted and limited its use and actionability for novices, such as in
the classroom or for class projects. To address these limitations, we crafted
design worksheets with steps to assist students walking through the
visualization design process for the first time. We validate the use of these
worksheets with 13 students in a graduate-level visualization course. In this
paper, we present the design activity worksheets --- concrete steps for students
to walk through the process of designing visualizations --- along with an
initial evaluation of the effectiveness of these worksheets, highlighting their
strengths and limitations.

\section{Related Work}

For the past few decades, pedagogy for data visualization and human-computer
interaction has begun to shift from academic or theoretical foundations toward
including skills for design, critique, and critical
analysis~\cite{Reimer2003,Rushmeier2007,Hearst2016b,He2017}. Educators have come
to realize that they must rapidly adapt their teaching methods to the growing
body of diverse students~\cite{Rheingans2016,Domik2016}, from flipped
classrooms~\cite{Rheingans2016,He2017} to online environments~\cite{Beyer2016}.
A recent approach among educators is to employ active
learning~\cite{Beyer2016,Godwin2016,Hearst2016b}, where techniques and methods
are used to encourage deeper analysis and synthesis as opposed to just passively
observing a lecture~\cite{Beyer2016}. Active learning can help overcome some of
the challenges faced by educators when teaching concepts surrounding design
thinking~\cite{Cennamo2011,He2017}, such as creating divergent
ideas~\cite{Roberts2015}. For example, VizItCards~\cite{He2017} was created to
help students practice and reinforce visualization concepts during workshops.
Other active learning approaches include the use of data visualization
exercises~\cite{Kerren2008}, rich discussions~\cite{Johnson2016,Craft2016},
design workshops~\cite{He2017,Huron2016,Zoss2016}, design
studios~\cite{Greenberg2009,Reimer2003}, and design games~\cite{Godwin2016}.

Within data visualization pedagogy, guidance for how to design data
visualizations is missing clear steps for novices. When teaching data
visualization design, educators often incorporate user interface
principles~\cite{Shneiderman2004}, taxonomies of data and
encoding~\cite{Munzner2014}, ideal visual principles~\cite{Tufte1986,cairo2012},
perceptual principles~\cite{Ware2010}, and generally empower students with the
ability to evaluate, criticize, and judge data visualizations. These concepts
are often then applied in the classroom through critiques or
projects~\cite{Eggermont2016}. Cumulative projects typically require students to
acquire their own datasets, come up with ideas to visualize data for different
tasks, and build an interactive, multiview visualization system. Students may
conduct their own design process according to design models provided in
textbooks~\cite{Ware2010,Shneiderman2004} or research
papers~\cite{Munzner2014,Sedlmair2012b}, but often these models are terminology
heavy, not actionable, and theoretical in nature. For students, it is
often useful to have a clear set of guidelines or instructions to walk through
this process for the first time. However, no such step-by-step guidance
currently exists for the data visualization design process.

Educators have worked on concretizing steps for the ideation process.
Specifically, the five design-sheet methodology~\cite{Roberts2015} utilizes
worksheets to structure and guide visualization students through the ideation
process. This approach by Roberts \etal was evaluated with 53 students as a way
to encourage engineering students to think divergently and creatively and sketch
out ideas on paper when first designing a visualization. However, in a workshop
at the 2016 IEEE VIS Conference that used these worksheets, we experienced a
limitation: many steps must occur first, such as data collection, identifying
the challenge, focusing on a target user, and finding tasks. Hence, in a
workshop or classroom setting, this ideation step often happens too soon in the
design process. Roberts \etal elude to these limitations as preparation
steps~\cite{Roberts2015}, but these steps can be nontrivial and tricky for
students. Thus, it would be beneficial to establish and evaluate more worksheets
beyond just ideation for data visualization design pedagogy.

\section{Worksheets for the Design Activity Framework}

The first contribution of this work is the creation of design worksheets that
follow the design process and decisions illustrated by the design activity
framework~\cite{McKenna2014}. Here, we discuss our process behind creating these
teaching materials and provide examples of their use. The worksheets and
teaching materials are located on a public-facing website\footnotemark for their
dissemination and use by others, and we encourage feedback and improvements to
these teaching materials by other visualization educators over time.

\footnotetext{Supplemental Materials can be accessed via:
\url{https://design-worksheets.github.io}}

Inspired by the five design-sheet methodology~\cite{Roberts2015}, we wanted to
integrate the visualization design process into visualization design worksheets
to enhance the teaching of a visualization design process. Our first goal was to
create a worksheet for each of the visualization design activities:
\textit{understand}, \textit{ideate}, \textit{make}, and \textit{deploy}. We
wanted to provide example methods and tips to enable students to go through a
visualization design process in its entirety. In order to create these
worksheets, we reflected on our combined 23 years of experience creating data
visualization tools and systems.

The primary author of this work coordinated the first draft of the worksheets,
by reflecting on his own experiences learning data visualization design in
course projects and applying this design knowledge across four design studies.
He was also inspired by the use of related worksheets by a colleague running
design studios in our university's architecture and design department. As a
result, he pinpointed distinct methods for generating and evaluating
visualization artifacts in each design activity. When identifying these methods,
we knew that engineering-type students could benefit from focusing on creating
many types of visualization artifacts, so we utilized the first four steps of
each activity for generation. For example, the \textit{ideate} sheet uses three
sketches for concept generation as in the five design-sheet
methodology~\cite{Roberts2015}.

We iterated upon the description of these steps and methods for each worksheet,
and we presented the worksheets to our research lab to acquire additional
feedback on their level of detail and utility. From this feedback, we received
recommendations to place more of a focus on the users earlier in the process and
to simplify complex, theoretical terminology, such as removing the use of the
nested model~\cite{Munzner2009,Meyer} in the original design. Thus, the methods
described on each worksheet were simplified and turned into a series of tangible
and generalized steps, as shown in Table~\ref{tab:steps}. We also walked through
a previous project~\cite{mckenna2016} using the worksheets to identify further
elements to add: more helper text, warning icons, expected results for each
step, and a label at the top for attachments. Lastly, we created several
introductory and template guides to help students fill out each worksheet, and
we include these resources in the Supplemental Materials.

\begin{table*}[htb]
  \centering
  \begin{tabular}[]{llll}
    \textit{\underline{understand}} & \textit{\underline{ideate}} & \textit{\underline{make}} & \textit{\underline{deploy}} \\
      identify the challenge \& users &
      select a design requirement &
      set an achievable goal &
      pinpoint a target audience \\
      find questions \& tasks &
      sketch first idea &
      plan encodings \& layouts &
      fix usability concerns \\
      check with users or explore data &
      sketch another idea &
      plan support for interactions &
      improve points of integration \\
      brainstorm design requirements &
      sketch final idea &
      sketching additional views &
      refine the aesthetics \\
      compare \& rank design requirements &
      compare \& relate your ideas &
      build the prototype \& check-in &
      consider a method to evaluate\\
  \end{tabular}\bigskip
  \caption{
    \label{tab:steps}
    Five steps for each design activity.
    We break down each visualization design activity~\cite{McKenna2014} into
    five concrete steps. The first four steps of each activity are generative,
    to establish design requirements, encoding and interaction sketches,
    visualization prototypes, or visualization systems. The fifth step is always
    evaluative, to compare different visualization artifacts in order to justify
    design decisions and record that reasoning down for later use. We shared
    these five steps with novice visualization designers, students, using design
    worksheets as a template, as in Figure~\ref{fig:worksheets}.
  }
\end{table*}

\begin{figure}
  \centering
  \fbox{\includegraphics[width=0.9\linewidth]{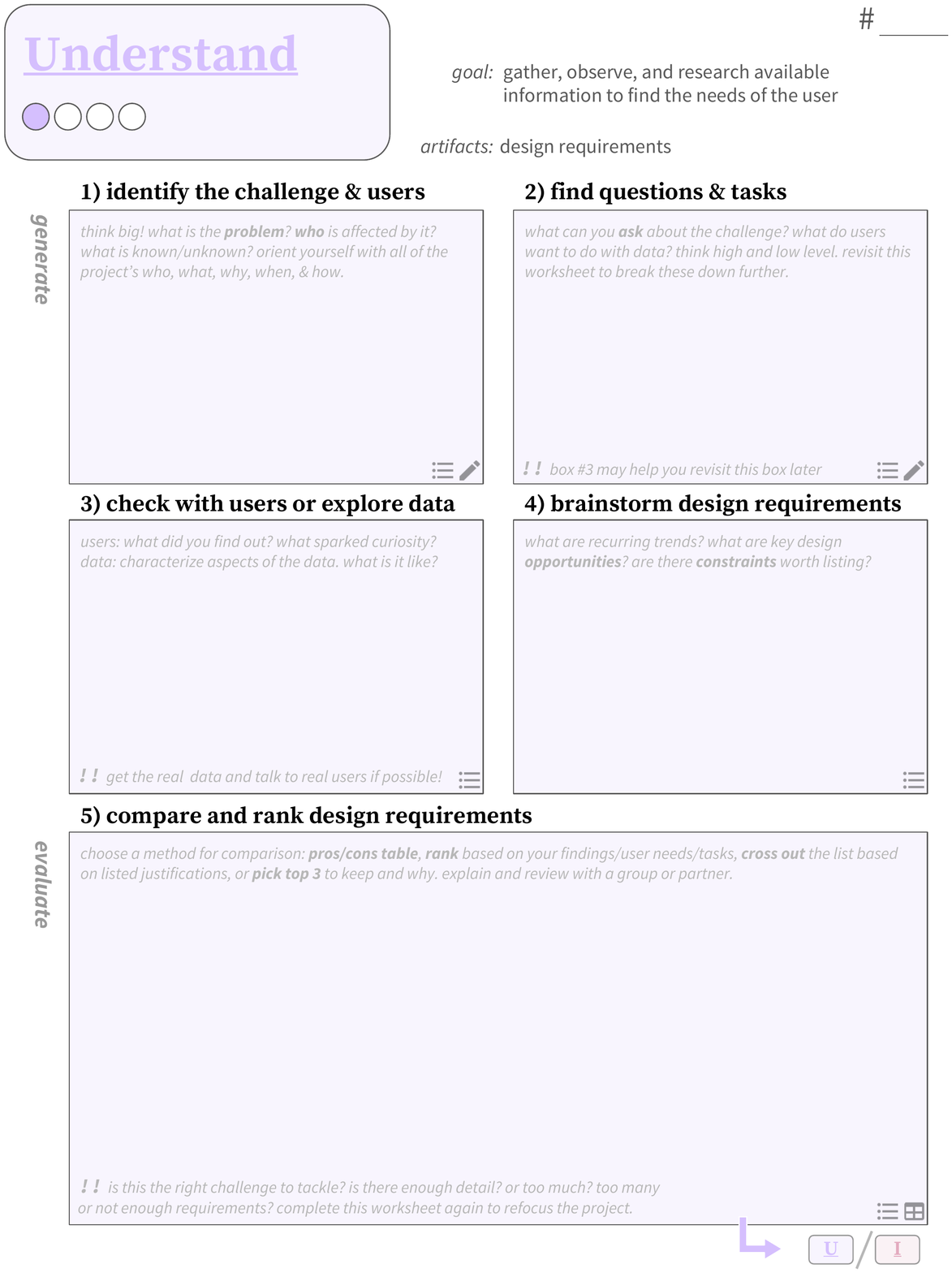}}
  \caption{
    \label{fig:worksheets}
    Worksheet for the \emph{understand} activity. We tailored this
    worksheet to help students identify their problem, users, data, and
    requirements for a data visualization system. Full page printouts of all
    four worksheets are included in the Supplemental Materials.
  }
\end{figure}

We introduce the first design worksheet for the \textit{understand} activity in
Figure~\ref{fig:worksheets}. At the top of each sheet, we describe the desired
goal and resulting visualization artifact or outcome for the activity. Each
sheet can be numbered in the top-right for keeping track of order and for
planning and retrospection. For each worksheet's five steps, we included
additional text to help students find the answer and complete each box. We added
warnings about when to jump back to previous boxes or worksheets, and icons to
illustrate the expected type of answer for each box: a list, a sketch, or a
table. Lastly, the bottom-right boxes point to the next potential activities of
the visualization design process.

\section{Evaluation Methodology}

Before introducing design worksheets to students, we needed to form a basis of
understanding, both in terminology and contextualized as a real-world
visualization example. We created an 80-minute lecture on visualization design
to teach the design activity framework~\cite{McKenna2014} to 66 students in our
university's graduate-level visualization course. This model was used to help
categorize visualization artifacts and design decisions that were contextualized
within a visualization design project, a cyber security visualization
dashboard~\cite{mckenna2016}. By utilizing this design study, we were able to
explain the design process with actual, tangible concepts.

\begin{figure}[htb]
  \centering
  \includegraphics[width=\linewidth]{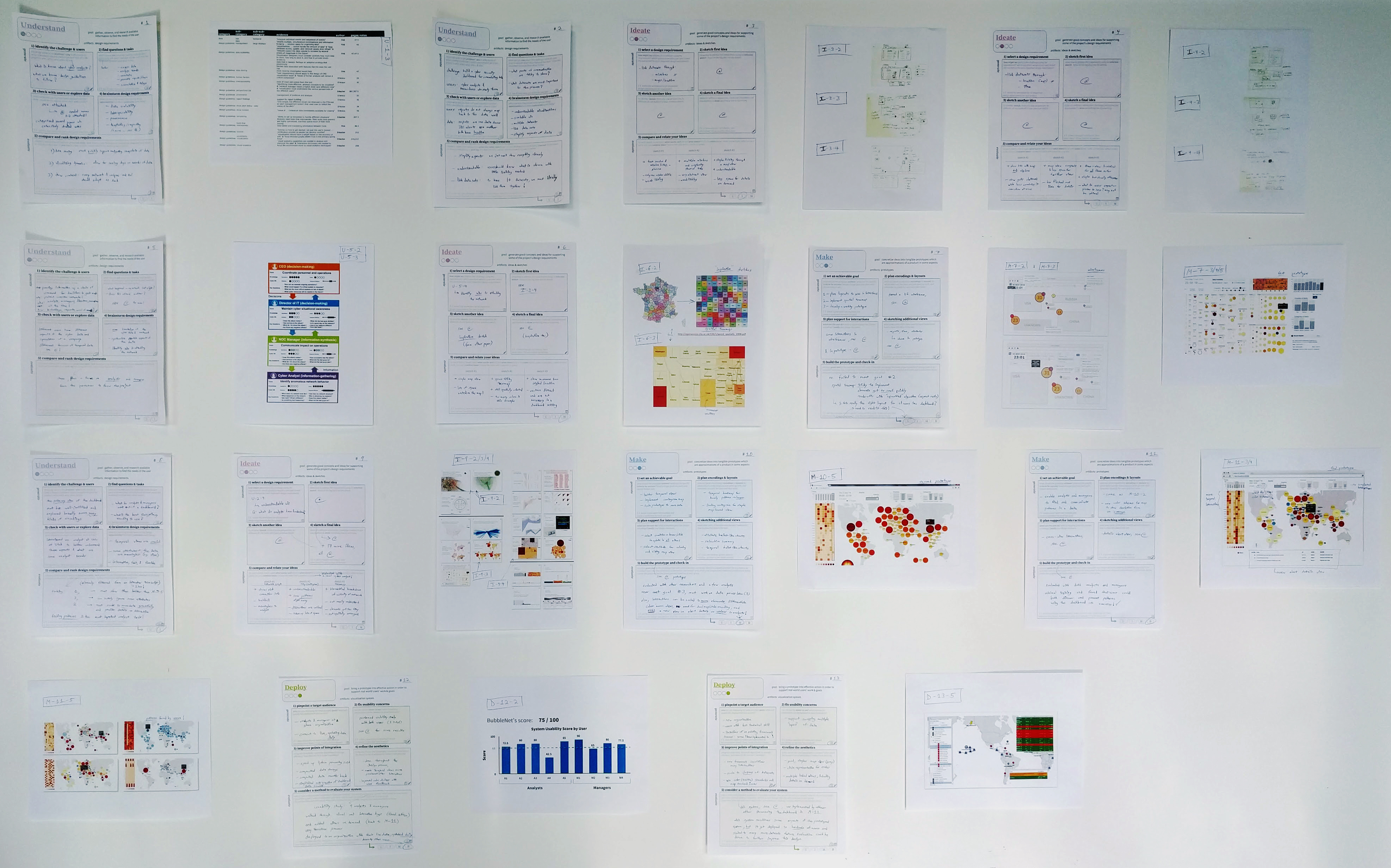}
  \caption{
    \label{fig:example}
    Design worksheet examples.
    We created example worksheets using linked sketches from our BubbleNet
    dashboard project~\cite{mckenna2016}. With this example, we taught students
    visualization design and showcased how a real-world, iterative design
    process can be captured using the worksheets. A detailed copy of each
    worksheet and all sketches are included in the Supplemental Materials.
  }
\end{figure}

The lecture was followed by an in-class exercise that had students analyze and
redesign an existing visualization using paper copies of the first two
worksheets: \textit{understand} and \textit{ideate}. Additionally, we mocked up
an example of how to use the design worksheets using the design study mentioned
previously. An overview of this example is provided in Figure~\ref{fig:example}.
As part of the course, students formed groups to complete a cumulative project:
to design and build a web-based interactive visualization system. We recruited
13 volunteers from the course to complete the design worksheets for each of
their six group projects, mentored by the primary author of this work. We also
include a copy of all our teaching resources and project details in the
Supplemental Materials.

The goal of this evaluation was to gain qualitative feedback on the worksheets
rather than compare pedagogical effectiveness; however, we recognize a
limitation of this methodology without a control group. We also had a limited
number of student volunteers and time availability from the primary author, plus
smaller group sizes are more manageable for qualitative methodologies. We were
also striving to preserve fairness in the classroom, so all students had equal
access to the volunteer opportunities, resources, and worksheets. For the
cumulative projects, we provided the worksheets in both paper and digital form
to all students, but only the 13 volunteers were required to submit digital or
scanned copies of their worksheets as part of their project. One student was not
part of the original volunteers, but due to complications with her project she
reached out to the teaching staff for further help and guidance for
visualization design within the context of her project.

To evaluate the efficacy of the worksheets in supporting the visualization
design process, we collected data from three sources: a full-course survey, a
focused survey on the worksheets, and interviews with student participants to
elicit in-depth worksheet feedback. Additionally, the mentor met weekly with
each group to provide feedback on their design process and on the worksheets.
These meetings provided a basis for obtaining in-person observations, in
addition to the feedback acquired anonymously through the surveys and detailed
interviews. The questions and prompts we utilized are included in the
Supplemental Materials.

The full-course survey was designed to gather anonymous feedback and assess the
general utility of the design worksheets for teaching the course. Specifically,
we asked questions about students' comfort level with design before and after
taking the course along with which factors taught them how to design
visualizations: lectures, in-class exercises, design worksheets, and the
cumulative project. In the focused survey for those who used the design
worksheets, we asked which worksheets worked well and which ones did not, and
why, along with 10 questions about the usefulness of the worksheets. To avoid
positivity bias, these questions varied between positive and negative wording.

After the student projects were completed, 11 students, at least one from each
project, participated in a semistructured interview to provide feedback on the
visualization design worksheets. The interview questions focused on digging
deeper into the survey findings. We asked open-ended questions to gather
suggestions for improvement. Lastly, we asked students to briefly describe steps
of the visualization design process in their own words.

\section{Evaluation Results}

For the full-class survey, we received 25 responses. Twenty-three students
showed an improvement in their comfort level for visualization design, on
average 2 out of 5 points higher by the end of the course. Students ranked these
improvements based on where they learned how to design, which was primarily
through the lectures, projects, and class exercises. The design worksheets
received a significantly larger portion of neutral responses for helping
students learn, possibly because only some students used them in their projects.
We compared the ratio of agreement to disagreement of these materials helping
students learn. The design worksheets were on the level of other methods
utilized in the course: design worksheets (13:1), lectures (23:1), exercises
(20:2), and projects (18:2).

For the survey sent to the students who used the visualization worksheets, we
received a total of seven responses. Overall, \textit{ideate} and
\textit{understand} worksheets were selected (six and four students,
respectively) as the most helpful design worksheets. Students stated that the
\textit{ideate} worksheet helped them critique their own designs, and
\textit{understand} helped jumpstart a visualization project. As stated by
students, \textit{ideate} \textit{``is the most clear worksheet''} and
\textit{``critique of one's own design was most helpful''} and that both
\textit{understand} and \textit{ideate} worksheets \textit{``helped to get the
project off the ground.''} On the flip side, the \textit{deploy} worksheet was
selected (four students) as the least helpful, because students often did not have
sufficient time to focus on this activity. Additional feedback highlighted some
drawbacks to the worksheets, such as vague terminology or phrasing, creative
limitations, and not enough structure. To uncover more information, we conducted
interviews as a follow-up.

During the follow-up interviews, all students described the design process using
elements of the visualization design activities. Specifically, four students
correctly recalled the names of each visualization design activity, but four
other students were unable to recall the \textit{deploy} activity --- possibly
since most groups did not go through in this activity. As in the survey, all
students found the \textit{understand} and \textit{ideate} worksheets the most
useful since they forced them to consider different tasks, users, and ideas.
Students noted that the worksheets provided a structured way to organize and
compare notes about different visualization design artifacts. Three students
stated that the worksheet example visualization project was helpful in
illustrating how to use the design worksheets. Nine students followed a design
process informed or exactly prescribed by the worksheets. One group acknowledged
that their design process, while different, still adhered to the steps provided
in each visualization design activity. Another student recognized the
flexibility of the worksheets: \textit{``If I had a different project, I would
use each box in different ways depending on the context''} (participant \#8, or
P8).

All students agreed that evaluation was a necessary and important step for
visualization design in order to pinpoint flaws in their understanding of the
problem, users, tasks, interactions, and encodings. Through evaluation, one
group discovered that their project was better suited to a subset of users, and
another group realized that a particular encoding resulted in points
overplotting. All students agreed that worksheets helped them document and
organize their visualization design process. These design worksheets served as a
\textit{``snapshot in time''} (P1) and were sufficiently detailed to explain
their design process for the final project report. Eight students described an
iterative process that occurred, although informal and not written on any of
their worksheets. Furthermore, the activities helped guide students as novice
designers, such as one student who used the visualization design worksheets for
the first time later in the course of the project and stated that \textit{``When
I used [the] worksheets it kept me focused on what I was doing and trying to get
more ideas or more [encodings]''} (P8).

An intriguing finding was that four students employed the worksheets in
surprisingly creative ways. For example, one student loaded the \textit{ideate}
worksheet in PDF form on her tablet and zoomed in to sketch various aspects of
her visualization design, allowing her to expand and use more space for the
visualization sketches. Also, another detail-oriented student transferred the
design worksheets into textual form, listing all of the steps and hints, so that
he could brainstorm and add detail to the problem and requirements over time, as
a living document. Four students expressed frustration with the paper worksheets
because they preferred another format, whether digital, larger paper, or the
ability to structure their notes how they wish. As one student put it,
\textit{``I think the concepts are very helpful in the worksheets .... [but] for
a free form thinker ... if you box it in then it is sort of restricting your
creativity, as it tells you how much you have to fit into where''} (P9).
Students suggested improvements and other feedback, which we explore next.

\section{Discussion}

To address these restrictions mentioned by that last student, a key improvement
recommended by five students was to convert the worksheets into a step-by-step
list, the same as the steps shown in Table~\ref{tab:steps}. Based on the
interviews, we recommend two formats for guiding the visualization design
process: a list and worksheets. The worksheets provide structure, \textit{``it's
like a checklist to make sure everything is covered''} (P11), but they also
limited free-form thinkers: \textit{``if you have a lot of things on your mind,
you won't fit everything in the box anyways''} (P6). Some visualization
designers recommend paper for sketching~\cite{Roberts2015}, but others in the
design community argue digital sketching can have functional benefits, such as
shapes, undo, layers, duplication, and manipulation of details through
zooming~\cite{Wu2011}, which two students utilized and felt was vital to their
visualization design process. An intriguing suggestion was to transform the
worksheets into an app: \textit{``a clickable, interactive worksheet, where you
click on this [and] it will connect you with the other worksheet and have a
screenshot''} (P8).

Students also suggested adding more worksheets to the materials. Six students
felt that \textit{``those [four] activities frame the process well''} (P2).
However, two students brought up a crucial aspect of evaluation and feedback:
that it might be worthwhile to devote a whole worksheet for evaluation,
otherwise \textit{``If you have it on the other worksheets, [evaluation] doesn't
seem to have as much value''} (P10). Four students requested a visualization
design worksheet to help probe into and explore the dataset or datasets that a
group may want to visualize. By providing guidance, steps, and questions on
aspects of the dataset, potential issues with visualizing the dataset could be
avoided later, and such issues occurred in three student projects. Lastly, three
students requested a visualization design worksheet on how to structure the code
of a visualization system, particularly in the case of one group with no
computer science background. Such a resource would help students brainstorm on
how to structure classes in their code, especially for building data
visualization systems.

Furthermore, some minor tweaks can be made to improve the visualization design
worksheets. Five students noted that having a good student example of the
worksheets would have helped define clear expectations for their work. We also
received recommendations to use a date-field rather than a blank number-field to
encourage students to simply organize their group worksheets over time as the
numbers were not often used and harder to coordinate among group members.
Students also suggested that each worksheet use a date-field rather than a blank
number-field to more easily coordinate work as a team. Most students did not
understand or use the helper text, result icons, and warning tips, so these
should be clarified and revisited in future work.

Nevertheless, the design worksheets helped guide students through actionable
steps for visualization design and facilitated effective discussions both within
a group and with their mentor. As students highlighted: \textit{``you break down
the process into those clear steps... an intuitive flow''} (P2), and:
\textit{``this was really good guidance for us ... well categorized for the
beginner''} (P3), and: \textit{``I didn't know where to start. It was nice to
have steps along the way''} (P4), and the benefit of generating ideas:
\textit{``we considered more options than we would have''} (P1). Despite the
many improvements that can be made, we emphasize that the worksheets provided
benefit to students, and a variety of future work could measure this success
more rigorously and compare how usable and effective these worksheets are for
students learning the visualization design process.

\section{Conclusion \& Future Work}

In this paper, we have introduced design worksheets to guide and teach novices
the process of designing a visualization system. These worksheets were designed
to simplify the theoretical concepts of the design activity
framework~\cite{McKenna2014}. We include all of the materials we used to teach
these concepts to 66 students in a graduate-level visualization course. We
evaluated the use of the design activities and worksheets through surveys and
interviews with 13 students. The results highlight what worked well and what
could be improved on these design worksheets going forward. Lastly, we
summarized these improvements and areas for future work on teaching
visualization design to novices.

These design worksheets are one possible step toward building more effective
teaching tools for data visualization and design, but plenty of work lies ahead.
One clear area for future work involves materials for design inspiration: from
visualization encodings to abstractions to tasks. Initial work shared by He and
Adar in Vizit Cards~\cite{He2017} is a step in this direction, and we encourage
the community to continue this line of work. While one student used VizIt cards,
she would have liked to see the cards generalized for other visualization
challenges. Furthermore, the design process steps and guidance can always be
improved to be more descriptive, more clear, sufficiently succinct, and
encompass other design methods and methodologies. Other common methods for
teaching are design studios~\cite{He2017} and exercises~\cite{Bertini2017}, and
it would be worthwhile to adapt design worksheets for these settings.

\acknowledgments{
  We thank the Visualization Design Lab and Professor James Agutter at the
  University of Utah for their feedback on this work. We would also like to
  thank the students of the Data Visualization course (Fall 2016), especially
  the volunteers who worked with us on the worksheets, without whom this work
  would not have been possible. This work is sponsored in part by the Air Force
  Research Laboratory and the DARPA XDATA program.
}


\bibliographystyle{styles/abbrv-doi}

\bibliography{references}

\begin{thebibliography}{10}

\bibitem{Bertini2017}
E.~Bertini.
\newblock {Teaching | Information Visualization}.
\newblock \url{http://enrico.bertini.io/teaching/}, 2017.
\newblock Accessed: 2017-02-03.

\bibitem{Beyer2016}
J.~Beyer, H.~Strobelt, M.~Oppermann, L.~Deslauriers, and H.~Pfister.
\newblock Teaching visualization for large and diverse classes on campus and
  online.
\newblock In {\em Pedagogy Data Visualization, IEEE VIS Workshop}, 2016.

\bibitem{cairo2012}
A.~Cairo.
\newblock {\em The Functional Art: An introduction to information graphics and
  visualization}.
\newblock New Riders, 2012.

\bibitem{Cennamo2011}
K.~Cennamo, S.~A. Douglas, M.~Vernon, C.~Brandt, B.~Scott, Y.~Reimer, and
  M.~McGrath.
\newblock {Promoting creativity in the computer science design studio}.
\newblock In {\em Proceedings of the 42nd ACM technical symposium on Computer
  science education - SIGCSE '11}, p. 649. ACM Press, New York, New York, USA,
  2011. doi: {{%
10\hspace{.1pt}\discretionary{.}{%
}{.}\hspace{.4pt}1145\discretionary{/}{%
}{/}1953163\hspace{.1pt}\discretionary{.}{%
}{.}\hspace{.4pt}1953344}}


\bibitem{Craft2016}
B.~Craft, R.-m. Emerson, and T.~J. Scott.
\newblock Using pedagogic design patterns for teaching and learning information
  visualization.
\newblock In {\em Pedagogy Data Visualization, IEEE VIS Workshop}, 2016.

\bibitem{Domik2016}
G.~Domik.
\newblock A data visualization course at the university of paderborn.
\newblock In {\em Pedagogy Data Visualization, IEEE VIS Workshop}, 2016.

\bibitem{Eggermont2016}
M.~Eggermont, C.~Perin, B.~Aseniero, and R.~Fallah.
\newblock Leveraging biological inspiration in an information visualization
  class.
\newblock In {\em Pedagogy Data Visualization, IEEE VIS Workshop}, 2016.

\bibitem{Godwin2016}
A.~Godwin.
\newblock Let's play: Design games and other strategies for introducing
  visualization through active learning.
\newblock In {\em Pedagogy Data Visualization, IEEE VIS Workshop}, 2016.

\bibitem{Greenberg2009}
S.~Greenberg.
\newblock {Embedding a design studio course in a conventional computer science
  program}.
\newblock In {\em Creativity and HCI: From experience to design in education},
  pp. 23--41. Springer, Boston, MA, 2009. doi: {{%
10\hspace{.1pt}\discretionary{.}{%
}{.}\hspace{.4pt}1007\discretionary{/}{%
}{/}978\discretionary{%
}{-}{-}0\discretionary{%
}{-}{-}387\discretionary{%
}{-}{-}89022\discretionary{%
}{-}{-}7\_3}}


\bibitem{He2017}
S.~He and E.~Adar.
\newblock {VizIt Cards: A card-based toolkit for infovis design education}.
\newblock {\em IEEE Transactions on Visualization and Computer Graphics}, 2017.
  doi: {{%
10\hspace{.1pt}\discretionary{.}{%
}{.}\hspace{.4pt}2450\discretionary{/}{%
}{/}2013\hspace{.1pt}\discretionary{.}{%
}{.}\hspace{.4pt}0043\discretionary{%
}{-}{-}13}}


\bibitem{Hearst2016b}
M.~A. Hearst.
\newblock Active learning assignments for student acquisition of design
  principles.
\newblock In {\em Pedagogy Data Visualization, IEEE VIS Workshop}, 2016.

\bibitem{Huron2016}
S.~Huron, S.~Carpendale, J.~Boy, and J.~D. Fekete.
\newblock Using {VisKit}: A manual for running a constructive visualization
  workshop.
\newblock In {\em Pedagogy Data Visualization, IEEE VIS Workshop}, 2016.

\bibitem{Johnson2016}
A.~Johnson.
\newblock Teaching data visualization in evl's cyber-commons classroom.
\newblock In {\em Pedagogy Data Visualization, IEEE VIS Workshop}, 2016.

\bibitem{Kerren2008}
A.~Kerren, J.~T. Stasko, and J.~Dykes.
\newblock {Teaching Information Visualization}.
\newblock In {\em Information Visualization}, pp. 65--91. Springer Berlin
  Heidelberg, Berlin, Heidelberg, 2008. doi: {{%
10\hspace{.1pt}\discretionary{.}{%
}{.}\hspace{.4pt}1007\discretionary{/}{%
}{/}978\discretionary{%
}{-}{-}3\discretionary{%
}{-}{-}540\discretionary{%
}{-}{-}70956\discretionary{%
}{-}{-}5\_4}}


\bibitem{McKenna2014}
S.~McKenna, D.~Mazur, J.~Agutter, and M.~Meyer.
\newblock {Design activity framework for visualization design}.
\newblock {\em IEEE Transactions on Visualization and Computer Graphics},
  20(12):2191--2200, 2014. doi: {{%
10\hspace{.1pt}\discretionary{.}{%
}{.}\hspace{.4pt}1109\discretionary{/}{%
}{/}TVCG\hspace{.1pt}\discretionary{.}{%
}{.}\hspace{.4pt}2014\hspace{.1pt}\discretionary{.}{%
}{.}\hspace{.4pt}2346331}}


\bibitem{mckenna2016}
S.~McKenna, D.~Staheli, C.~Fulcher, and M.~Meyer.
\newblock Bubblenet: A cyber security dashboard for visualizing patterns.
\newblock {\em Eurographics Conference on Visualization (EuroVis)},
  (just-accepted), 2016.

\bibitem{Meyer}
M.~Meyer, M.~Sedlmair, P.~S. Quinan, and T.~Munzner.
\newblock {The nested blocks and guidelines model}.
\newblock {\em Information Visualization}, 2013.

\bibitem{Munzner2009}
T.~Munzner.
\newblock {A nested model for visualization design and validation}.
\newblock {\em IEEE Transactions on Visualization and Computer Graphics},
  15(6):921--928, 2009. doi: {{%
10\hspace{.1pt}\discretionary{.}{%
}{.}\hspace{.4pt}1109\discretionary{/}{%
}{/}TVCG\hspace{.1pt}\discretionary{.}{%
}{.}\hspace{.4pt}2009\hspace{.1pt}\discretionary{.}{%
}{.}\hspace{.4pt}111}}


\bibitem{Munzner2014}
T.~Munzner.
\newblock {\em {Visualization Analysis and Design}}.
\newblock CRC Press, 2014.

\bibitem{Reimer2003}
Y.~J. Reimer and S.~A. Douglas.
\newblock {Teaching HCI design with the studio approach}.
\newblock {\em Computer Science Education}, 13(3):191--205, sep 2003. doi: {{%
10\hspace{.1pt}\discretionary{.}{%
}{.}\hspace{.4pt}1076\discretionary{/}{%
}{/}csed\hspace{.1pt}\discretionary{.}{%
}{.}\hspace{.4pt}13\hspace{.1pt}\discretionary{.}{%
}{.}\hspace{.4pt}3\hspace{.1pt}\discretionary{.}{%
}{.}\hspace{.4pt}191\hspace{.1pt}\discretionary{.}{%
}{.}\hspace{.4pt}14945}}


\bibitem{Rheingans2016}
P.~Rheingans.
\newblock Minor adventures in flipped classrooms, team-based learning, and
  other pedagogical buzzwords.
\newblock In {\em Pedagogy Data Visualization, IEEE VIS Workshop}, 2016.

\bibitem{Roberts2015}
J.~Roberts, C.~Headleand, and P.~Ritsos.
\newblock {Sketching designs using the five design-sheet methodology}.
\newblock {\em IEEE Transactions on Visualization and Computer Graphics},
  PP(99):1--1, 2015. doi: {{%
10\hspace{.1pt}\discretionary{.}{%
}{.}\hspace{.4pt}1109\discretionary{/}{%
}{/}TVCG\hspace{.1pt}\discretionary{.}{%
}{.}\hspace{.4pt}2015\hspace{.1pt}\discretionary{.}{%
}{.}\hspace{.4pt}2467271}}


\bibitem{Rushmeier2007}
H.~Rushmeier, J.~Dykes, J.~Dill, and P.~Yoon.
\newblock {Revisiting the need for formal education in visualization}.
\newblock {\em IEEE Computer Graphics and Applications}, 27(6):12--16, nov
  2007. doi: {{%
10\hspace{.1pt}\discretionary{.}{%
}{.}\hspace{.4pt}1109\discretionary{/}{%
}{/}MCG\hspace{.1pt}\discretionary{.}{%
}{.}\hspace{.4pt}2007\hspace{.1pt}\discretionary{.}{%
}{.}\hspace{.4pt}156}}


\bibitem{Sedlmair2012b}
M.~Sedlmair, M.~Meyer, and T.~Munzner.
\newblock {Design study methodology: Reflections from the trenches and the
  stacks}.
\newblock {\em IEEE Transactions on Visualization and Computer Graphics},
  18(12):2431--2440, 2012.

\bibitem{Shneiderman2004}
B.~Shneiderman and C.~Plaisant.
\newblock {\em {Designing the User Interface : Strategies for Effective
  Human-Computer Interaction}}.
\newblock 2004.

\bibitem{Tufte1986}
E.~R. Tufte.
\newblock {\em The Visual Display of Quantitative Information}.
\newblock Graphics Press, Cheshire, CT, USA, 1986.

\bibitem{Ware2010}
C.~Ware.
\newblock {\em {Visual Thinking: for Design}}.
\newblock Morgan Kaufmann, 2010.

\bibitem{Wu2011}
J.-C. Wu, C.-C. Chen, and H.-C. Chen.
\newblock {Comparison of designer's design thinking modes in digital and
  traditional sketches}.
\newblock {\em Design \& Technology Education}, 17(3):37--48, Nov. 2012.

\bibitem{Zoss2016}
A.~Zoss.
\newblock Challenges and solutions for short-form data visualization
  instruction.
\newblock In {\em Pedagogy Data Visualization, IEEE VIS Workshop}, 2016.

\end{thebibliography}


\end{document}